\documentclass[prl,onecolumn,showpacs,amsmath,amssymb]{revtex4-1}
\usepackage{graphicx}       
\usepackage{epstopdf}
\usepackage{dcolumn}        
\begin{document}
\title{Multi-band superconductivity and nanoscale inhomogeneity at oxide interfaces}
\author{S. Caprara$^{1,2}$, J. Biscaras$^3$, N. Bergeal$^3$, D. Bucheli$^1$, S. Hurand$^3$, C.
Feuillet-Palma$^3$, A. Rastogi$^4$, R. C. Budhani$^{4,5}$, 
J. Lesueur$^3$, and M. Grilli$^{1,2}$}
\affiliation{$^1$Dipartimento di Fisica, Universit\`a di Roma ``La Sapienza'', Piazzale Aldo Moro 5, 00185 Roma, 
Italy\\ 
$^2$ISC-CNR and Consorzio Nazionale Interuniversitario per le Scienze Fisiche della Materia, Unit\`a 
di Roma ``Sapienza''\\
$^3$LPEM-UMR8213/CNRS-ESPCI Paris Tech - UPMC, 10 rue Vauquelin - 75005 Paris, France\\
$^4$Condensed Matter - Low Dimensional Systems Laboratory, Department of Physics, Indian Institute of Technology Kanpur, Kanpur 208016, India \\
$^5$National Physical Laboratory, New Delhi - 110012, India}
\begin{abstract}
{The two-dimensional electron gas at the LaTiO$_3$/SrTiO$_3$ or LaAlO$_3$/SrTiO$_3$ oxide interfaces becomes 
superconducting when the carrier density is tuned by gating. The measured resistance and superfluid density reveal 
an inhomogeneous superconductivity resulting from percolation of filamentary structures of superconducting  ``puddles'' with randomly 
distributed critical temperatures, embedded in a non-superconducting matrix. Following the evidence that 
superconductivity is related to the appearance of high-mobility carriers, we model intra-puddle superconductivity by 
a multi-band system within a weak coupling BCS scheme. The microscopic parameters, extracted by fitting the 
transport data with a percolative model, yield a consistent description of the dependence of the average 
intra-puddle critical temperature and superfluid density on the carrier density.}
\end{abstract}
\date{\today}
\maketitle

The discovery of superconductivity in the two-dimensional electron gas (2DEG) formed at certain oxide interfaces such as 
 LaAlO$_3$/SrTiO$_3$ (LAO/STO) or LaTiO$_3$/SrTiO$_3$ (LTO/STO)
\cite{reyren,triscone,espci1,espci2} has attracted great interest and stimulated intense research activity. The possibility of tuning the carrier density by means of electrostatic 
gating in these extremely two-dimensional superconductors (the Fermi wavelength being smaller than the thickness of the gas\cite{espci2}) opens new opportunities to study 
fundamental issues in quantum fluids\cite{espci3}. More generally, these 2DEG appear as very rich quantum matter systems, since they also present magnetism 
\cite{ariando,luli,bert,bert2012}  and strong tunable spin-orbit coupling \cite{cavigliaprl,CPG}. As for every two dimensional systems, disorder is expected to play a relevant role 
on their physical properties.
Recent magnetotransport measurements revealed the existence of two kinds of carriers in LTO/STO, with high and low mobility respectively\cite{espci2}. 
It has been suggested 
that superconductivity is related to the appearance of the high-mobility carriers\cite{espci2,bell}
with an inhomogeneous character, i.e.  superconducting  ``puddles" embedded in a 
(weakly localizing) metallic background\cite{espci3}. Direct superfluid density\cite{bert2012} and magnetic\cite{luli} measurements display signs of strong inhomogeneity at 
micrometric scales. Recent current distribution\cite{Moller} and surface potential studies\cite{ilani2} reveal stripes structures in these interfaces. Given all these signs of 
inhomogeneities, one has first to analyze the role of disorder before getting further insight in the physics of the 2DEG, and for example, explain the rapid decrease of $T_c$ 
with the gate voltage while under-doping. 

Our aim is on the one hand to infer a more detailed structure of the inhomogeneous state and on the other hand to provide a description for superconductivity in an 
inhomogeneous system with different kinds of carriers. We first show that resistance measurements\cite{espci2} and the topographic mapping of the superfluid density on 
micrometric  scale\cite{bert2012} can be phenomenologically accounted for within a scheme of a percolative inhomogeneous 2DEG with poor long-distance connectivity, but 
substantial superconducting (SC) current loops to warrant a high fraction of diamagnetic response. 
We then show that the properties of the SC puddles (e.g., their fraction, and critical 
temperatures) can be extracted from experiments and used at a microscopic level to model the intra-puddle multi-carrier pairing, gaining insight about the pairing mechanism. 
Although some features of the diamagnetic response are seemingly related to a strong coupling SC behavior\cite{bert2012}, our analysis demonstrates that inhomogeneities
and multi-band superconductivity fully account for the SC behavior of these systems within a standard weak coupling BCS scheme.

{\it --- Sheet resistance ---} We have recently shown that the percolative SC transition in 
inhomogeneous systems, like the oxide interfaces, is well described by the Effective Medium Theory (EMT) \cite{CGBC}. 
For samples exhibiting a resistance drop due to superconductivity, the measured sheet resistance of LTO/STO is well fitted 
by the EMT with a SC fraction $w<1$ (the SC puddles). Each puddle has a random local 
critical temperature $T_c$, which we assume to have a Gaussian distribution, with mean $\overline T_c$ and width 
$\gamma$. The weight $w$ of the $T_c$ distribution represents the fraction of SC puddles and the 
non SC fraction $1-w$ represents the metallic background (we refer the reader to Ref. \cite{CGBC} for details). The sheet resistance at any temperature is
\[
R(T)=R^\infty\left[w\,\mathrm{erf}\,\left(\frac{T-\overline T_c}{\sqrt{2}\gamma}\right)
+1-w\right],
\]
with fitting parameters $R^\infty$ (high-temperature sheet resistance), $w$, $\overline T_c$, and 
$\gamma$. The percolative SC transition, marked by a vanishing $R$, takes place 
at a temperature $T_p\le \overline T_c$, such that 
\begin{equation}
\mathrm{erf}\,\left(\frac{T_p-\overline T_c}{\sqrt{2}\gamma}\right)=1-\frac{1}{w}.
\label{weight}
\end{equation}
This equation has a solution for $T_p$ only if $w\ge\frac{1}{2}$, i.e., if the SC fraction can 
percolate in the two-dimensional system. When $T_p$ becomes negative or is not defined, the system remains resistive 
down to $T=0$, although a significant reduction of sheet resistance may still occur, witnessing the presence of a 
sizable (though not percolating) SC fraction.

\begin{figure}
\vspace{-1 truecm}
\includegraphics[width=1.05\linewidth]{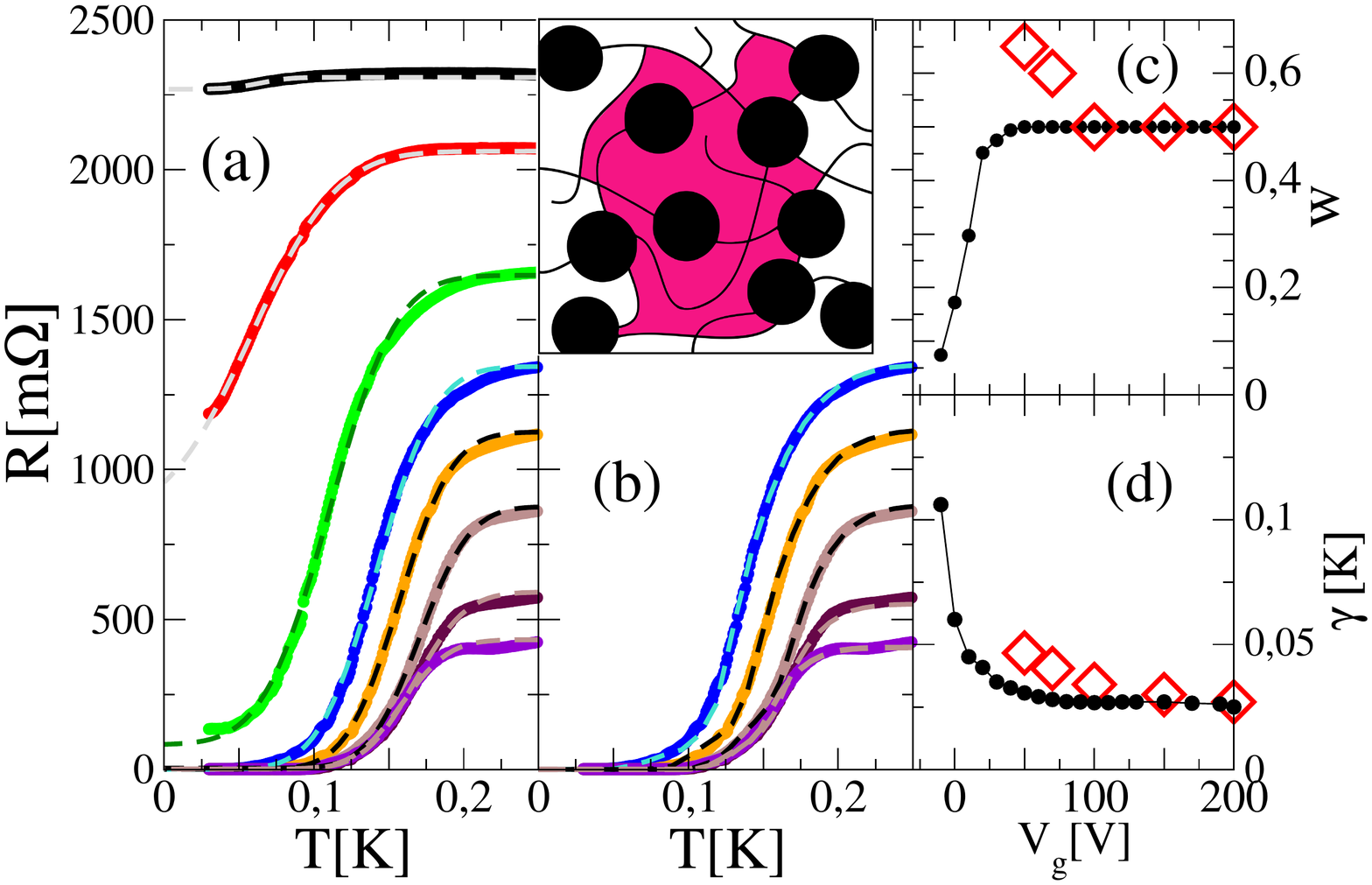}
\caption{(a) Sheet resistance as a function of temperature in a LTO/STO sample for various values of the gate voltage. 
 From top to bottom $V_g=-10$, $10$, $30$, $50$, $70$, $100$, $150$, $200$\,V.
Symbols mark the experimental data, while the dashed lines are the theoretical fits using the EMT (see text).
(b) From top to bottom $V_g=50$, $70$, $100$, $150$, $200$\,V.
Symbols mark the experimental data, while the dashed lines are the theoretical fits using the RRN (see text).
Inset: sketch of an inhomogeneous RRN system with a SC filamentary+baubles cluster. The shaded region 
is a closed loop contributing to the diamagnetic response.
(c) SC fraction $w$ extracted from the fits of the sheet resistance within
EMT (filled circles, solid line) and within the RRN (empty diamonds).
(d) Width of the Gaussian distribution of $T_c$ used in the fits of the sheet resistance
within the EMT (filled circles, solid line) and within the RRN  (empty diamonds).}
%
\label{fig-rho-vs-t}
\end{figure}

Fig. \ref{fig-rho-vs-t} (a) reports the measured sheet resistance as a function of temperature for various values 
of the gate voltage $V_g$ in a LTO/STO sample, and the fitting EMT curves. From the fits one can extract the average 
intra-puddle critical temperature $\overline T_c$ [corresponding to the temperature of maximal slope of 
$R(T)$ within EMT], which is reported in Fig. \ref{fig-Tc} (a) as a function of $V_g$ (red solid 
line and empty circles). The fits also provide the fraction of SC regions $w$ [displayed in Fig. 
\ref{fig-rho-vs-t} (c) as a function of $V_g$]. Since EMT disregards spatial correlations, the simultaneous 
request of a percolating SC cluster and the tailish shape of the resistance near percolation force the 
weight of the regions which can become SC to be $w\approx \frac{1}{2}$ \cite{CGBC} (Fig. \ref{fig-rho-vs-t}(c), see also Supplemental 
Material).  Fig. \ref{fig-rho-vs-t}(d) reports  
the width $\gamma$ of the Gaussian distribution of $T_c$, which diverges as the fraction $w$ of the SC puddles goes to zero. This is rather natural
because the density decrease emphasizes the effects of disorder so that
fluctuations of the local superconductivity increase, leading to a substantial broadening of the $T_c$ distribution. We will show later a more precise description of this phenomenon within the two-bands model.

{\it ---  Superfluid density ---}
The superfluid density was measured in LAO/STO interfaces using a scanning SQUID probe averaging over micrometric 
regions \cite{bert2012}. This technique is not sensitive to submicrometric inhomogeneities, but revealed a 
{\it distribution} of superfluid densities within a given sample on the micrometric scale, thereby supporting the idea 
of inhomogeneous 2DEG at these oxide interfaces. Since the sheet resistance curves of these systems are similar to 
those of LTO/STO interfaces, we  assume that our percolative EMT analysis also applies in this case and that 
the measured local superfluid density is an average over an inhomogeneous state of submicrometric puddles \cite{nota-submicron}.

\begin{figure}
\includegraphics[angle=0,scale=0.3]{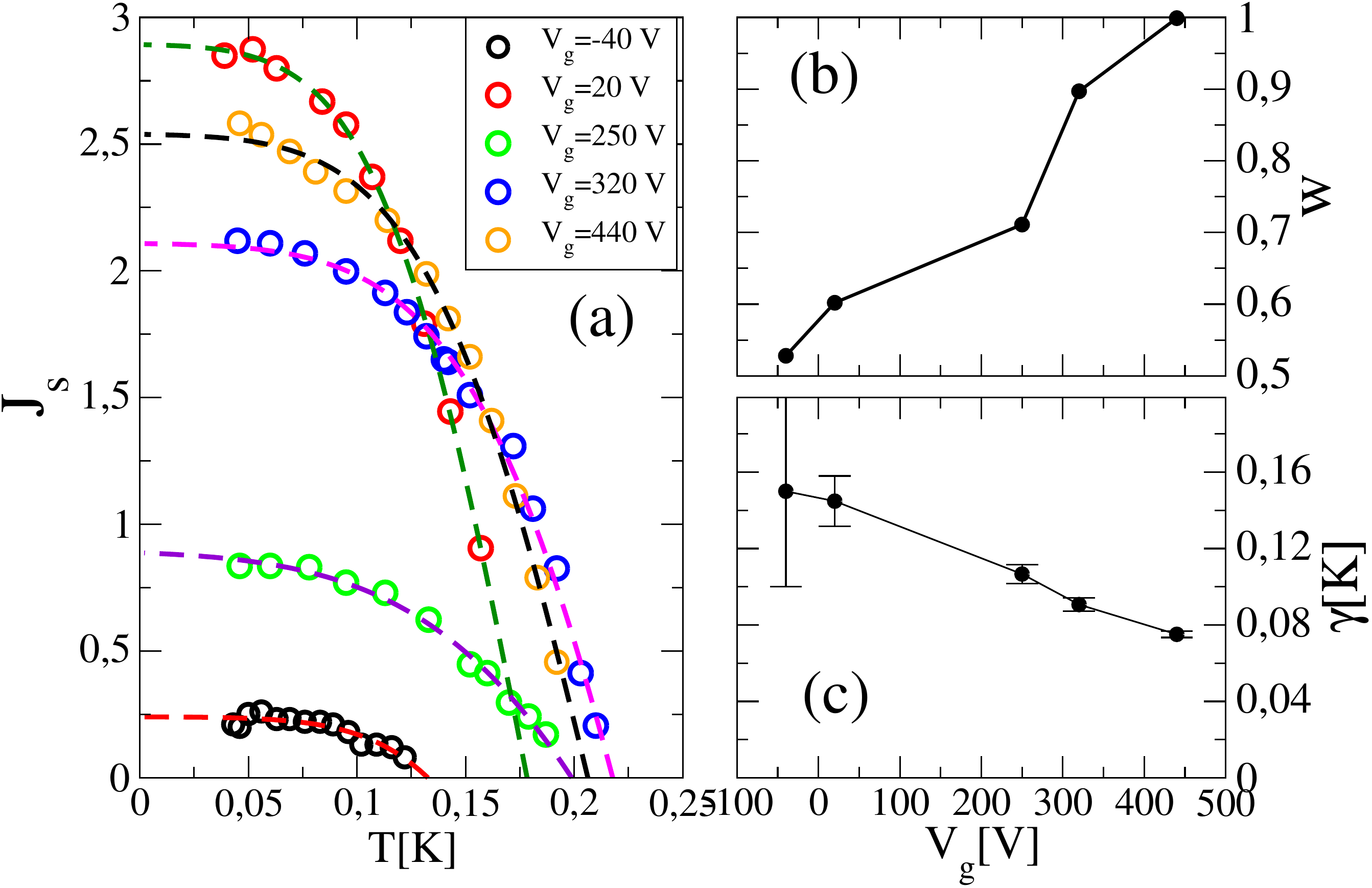}
\caption{(a) EMT fits of the superfluid density in LAO/STO (lines) as a function of temperature for various gate 
voltages. The experimental data (symbols) are extracted from Ref. \onlinecite{bert2012}. (b) SC fraction $w$
responsible for the diamagnetic response. (c) ) Width of the Gaussian distribution of $T_c$ used in the fits of the
superfluid density.}
%
\label{figstiff}
\end{figure}

\begin{figure}
\includegraphics[angle=0,scale=0.3]{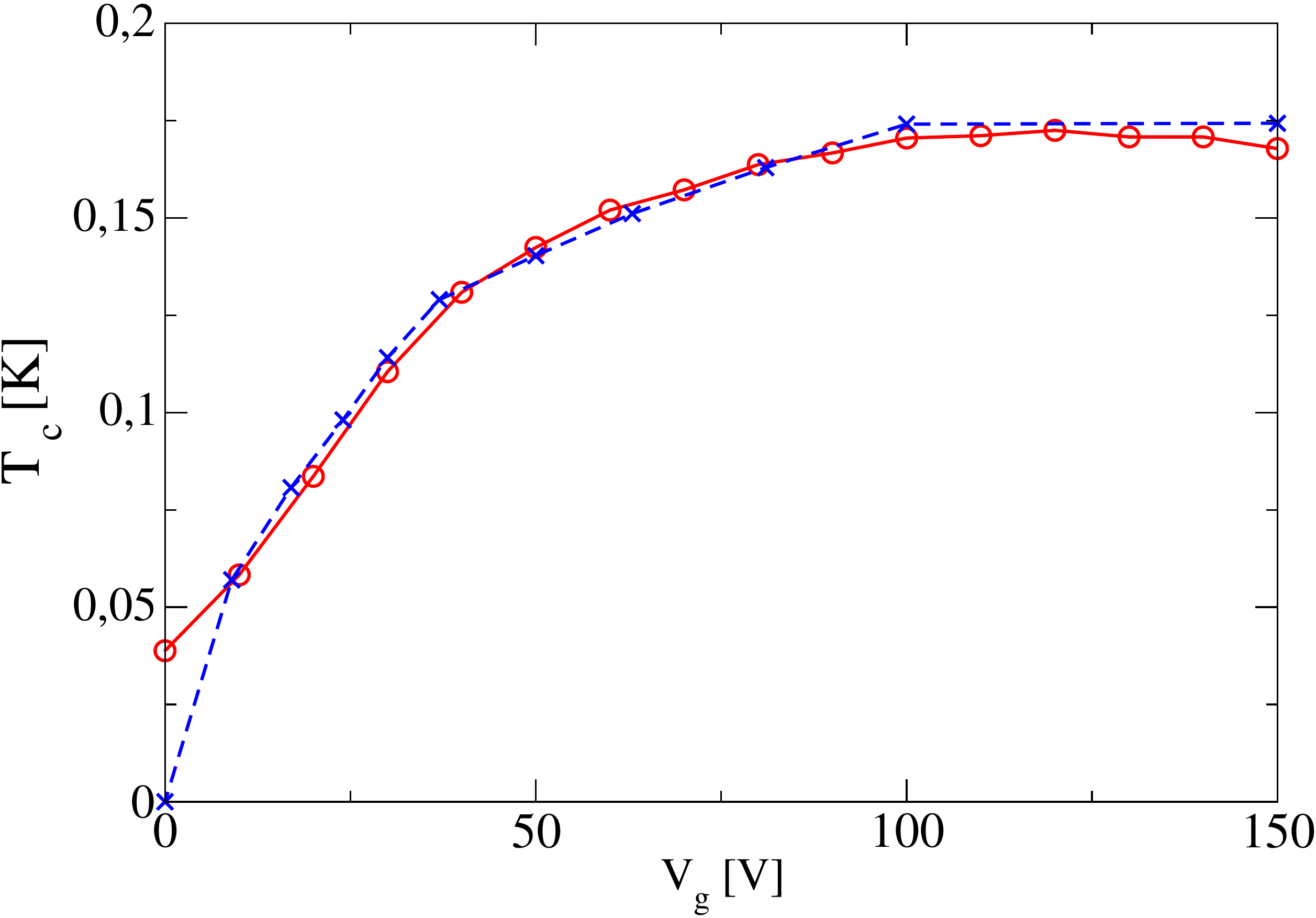}
\caption{ $\overline T_c$ extracted from the EMT fit to experimental data (red solid line and empty circles) and the 
theoretical
$T_c$ calculated within the two band model  (see text), as a function of the gate voltage $V_g$ 
(blue dashed line and crosses).
}
%
\label{fig-Tc}
\end{figure}

To capture this effect, we extend the EMT to small but finite frequency $\omega$. We model the metallic regions with 
a Drude-like complex conductivity $\sigma_N(\omega)=B/(A+i\omega)$ and the SC regions with a purely 
reactive conductivity $\sigma_S(\omega)=B/(i\omega)$. We introduce the quantities 
$\rho_S(\omega)=1/\sigma_S(\omega)=i\omega/B$ and $\rho_N(\omega)=1/\sigma_N(\omega)=\rho_0+\rho_S(\omega)$, with 
$\rho_0=A/B$. At high temperature, the system is metallic and $\rho(\omega)=1/\sigma(\omega)=\rho_N$ everywhere. 
However, within the SC puddles, $\rho_0$ vanishes as soon as the local critical temperature is reached. 
The solution of the EMT equation, neglecting higher frequency terms $\sim\omega^2$, is
\[
\rho(\omega)\approx\rho_0(w_N-w_S)\vartheta(w_N-w_S)+\frac{\rho_S(\omega)}{|w_N-w_S|},
\]
where $\vartheta(x)$ is the Heaviside step function, $w_S$ is the fraction of puddles that have become 
SC (at a given temperature $T$), and $w_N=1-w_S$ is the metallic fraction (resulting both from puddles 
that have not yet become SC and from the metallic background). Evidently, when $w_N>w_S$, the system displays 
a Drude-like complex conductivity. However, below the percolation temperature $T_p$ (if any), $w_S>w_N$, and we find 
a purely reactive conductivity
\[
\sigma(\omega)=\frac{1}{\rho(\omega)}=\frac{A(w_S-w_N)}{i\omega}.
\]
Therefore, for $T<T_p$ and $w$ given by Eq.(\ref{weight}), the superfluid density of the 
percolating network is
\begin{eqnarray}
J_s &\propto& w_S-w_N = w-1-w\,\mathrm{erf}\,\left(\frac{T-\overline T_c}{\sqrt{2}\gamma}\right)\nonumber\\
&=&w\left[ \mathrm{erf}\,\left(\frac{T_p-\overline T_c}{\sqrt{2}\gamma}\right)-
\mathrm{erf}\,\left(\frac{T-\overline T_c}{\sqrt{2}\gamma}\right)\right]. 
\label{stiffness}
\end{eqnarray}
Fig. \ref{figstiff} (a) reports this micrometrically averaged superfluid density together with our EMT fits
[Eq. (\ref{stiffness})] at various gate voltages $V_g$. Remarkably, the $T$ dependence of 
$J_s(T)/J_s(T=0)$ mimics the qualitative behavior of the BCS prediction, but may quantitatively deviate from it.
The slope at the transition, for instance, is ruled by the width $\gamma$ of the distribution of critical 
temperatures. 
We can thus account for the deviations from standard BCS observed in Ref. \onlinecite{bert2012}, which were attributed to 
a tendency to strong coupling. Our alternative explanation for these deviations is that the system is
in a weak coupling regime, but the percolative superfluid
 density is ruled by the submicrometric distribution of inhomogeneities averaged at the micron scale by the SQUID 
pick-up loop used in Ref \onlinecite{bert2012}. The fits allow to obtain the fraction $w$ of SC puddles extracted from diamagnetic measurements
 which is reported in Fig. \ref{figstiff}(b). It ranges from
1/2 to 1, and is therefore always larger than the fraction obtained
from transport measurements. This is fully consistent within a percolative description of the system: resistance
measurements mainly probe the direct percolative path, regardless of dead-ends and disconnected SC regions\cite{stauffer},
while screening measurements are sensitive to all SC loops, even when not connected to the backbone. In this case, 
the diamagnetic fraction can be nearly one, while the connectivity for macroscopic transport can be very
small (or even vanishing) if many SC puddles or loops are disconnected [see the sketch in the inset of Fig.  \ref{fig-rho-vs-t}(b)].
Fig. \ref{figstiff}(c) displays the behavior of the width $\gamma$ of the $T_c$ distribution. Despite the differences in material
and physical quantities, $\gamma$ has a similar behavior to the $\gamma$ from transport in LTO/STO [Fig. \ref{fig-rho-vs-t}(d)]
and is of comparable magnitude.

{\it --- Effect of space correlations ---} 
The explanation of the above discrepancy between the SC weight $w$ in transport and magnetic experiments rests on the
filamentary structure of the SC fraction in LXO/STO systems, which is one of the two main outcomes of our work. 
However, the above results (and fits) have been obtained within the EMT, which is a mean-field-like 
approach neglecting space correlations. Therefore, to substantiate the above interpretation,
it is important to test how robust is the finding that the SC fraction involved in transport is rather small, $w\approx 0.5$,
within a model, which takes into account the space correlations inherent in the filamentary structure.
To this purpose, we 
refine our model and solve a random resistor network (RRN) where the SC regions only occur on a 
cluster with strong space correlations embedded in a metallic matrix. Based on a preliminary analysis 
\cite{BCCG}, showing that the tailish sheet resistance implies both a dense SC cluster at short distances 
and a filamentary structure at larger distance, we numerically generate a fractal-like structure\cite{BCCG,notafractal}. 
Due to the poor long-distance connectivity, percolation only occurs when the resistance of
almost all bonds has been switched off. For a Gaussian distribution of critical temperatures, the system is thus 
forced to explore the low temperature asymptotic of the distribution, thereby producing a tailish resistance\cite{notabulky}. 
On the other hand, a purely filamentary structure, although dense at short distance due to the almost fractal 
cluster, is still too faint ($w\lesssim 0.3$) and fails to reproduce accurately the high temperature part of the $R(T)$ curves. Thus
in order to vary  the density of the SC cluster without crucially changing the 
long-range connectivity, we decorate the filaments with randomly distributed circular ``superpuddles'', their number 
being chosen to yield RRNs with weights $w$ ranging from $0.3$ to $0.7$ (see Supplemental Material). The resulting 
clusters look like a filamentary ``Christmas tree'' with decorating baubles [see inset of Fig. \ref{fig-rho-vs-t} (b)]. Fig. 
\ref{fig-rho-vs-t} (b) reports fits of the sheet resistance using resistivity curves calculated on such Christmas 
trees. Noticeably, even though the SC weight $w$ is not forced to $\frac{1}{2}$ like in EMT, we still 
find that the sheet resistance only acquires a tailish behavior if $0.5\lesssim w\lesssim 0.65$ (Fig.  \ref{fig-rho-vs-t} (c)), where the lower 
bound is imposed by the high slope at intermediate temperatures and the upper bound is due to the tailish behavior 
close to percolation. Fig. \ref{fig-rho-vs-t} (d) reports the width $\gamma$ of the Gaussian distribution of $T_c$ 
obtained within the Christmas tree model. One can see that $\gamma$ follows the same qualitative behavior as in the EMT case
with a substantial increase upon lowering the voltage. As a conclusion, EMT and RRN models lead to very close 
results about the $T_c$ distribution and its variation with the gate voltage provided a filamentary Christmas tree 
structure is used for the RRN.

{\it --- Multiband BCS scenario ---}
The inhomogeneous character of LXO/STO systems, clearly apparent both in the normal-state [$R(T)$] 
and in the SC [$J_s(T)$] properties, entails a distribution of local critical temperatures with an average depending
on the overall density (i.e., gating) $\overline T_c(V_g)$. This dependence can be extracted from the resistance fits
[Fig. \ref{fig-Tc}] and it provides insight on the 
the intra-puddle pairing mechanism. A recent magnetotransport analysis \cite{espci2} demonstrated the coexistence of a 
rather large density of low-mobility carriers with a smaller density of high-mobility carriers in the 
SC regime of LTO/STO. In particular, superconductivity seems to be triggered by the presence of the 
high-mobility carriers. A similar behavior seems to occur in LAO/STO \cite{Rakhmilevitch}. 
According to this scenario, we propose that the 2DEG formed at the LXO/STO interface may be described as a 
multi-band system \cite{salluzzo,ilani,delugas,held}, and the occurrence of superconductivity may be captured by means of a 
BCS-like Hamiltonian (see Supplemental Material). We take for the $\ell$-th sub-band of our 2DEG a parabolic 
dispersion law
\[
\varepsilon_{\mathbf{k},\ell}=\bar\varepsilon_\ell+\frac{\hbar^2k_x^2}{2 m_{\ell,x}}+
\frac{\hbar^2k_y^2}{2 m_{\ell,y}}, 
\]
where $m$ is the (possibly anisotropic) effective mass of the charge carriers. Following the evidence 
that superconductivity is related to the appearance of high-mobility carriers, we represent the whole set of 
low-lying bands with one sub-band ($\ell=1$) collecting all the low-mobility carriers with vanishingly small $T_c$, 
while the high-mobility carriers in the sub-band $\ell=2$ give rise to a finite $T_c$. Thus, SC puddles 
are regions where the sub-band $\ell=2$ is locally filled, whereas the (weakly localizing) metallic background 
corresponds to regions where the sub-band $\ell=2$ is empty. To reproduce the experimental results, we are led 
to a suitable choice of the pairing amplitudes 
$g_{\ell\ell'}$ (intraband for $\ell=\ell'$, interband, for $\ell\neq\ell'$): $g_{11}\ll (g_{12},g_{21})\ll g_{22}$ 
(this condition is consistent with the analysis of a two-band model in Ref. \onlinecite{balatsky}).
According to the standard BCS approach, the pairing amplitudes are only effective in a limited dynamical range 
$|\varepsilon_{\mathbf{k},\ell}-\mu|,|\varepsilon_{\mathbf{k}',\ell'}-\mu|\le \hbar\omega_0$, where $\omega_0$ is 
a characteristic frequency of the mode that mediates pairing and $\mu$ is the chemical potential.
We assume henceforth that the bottoms of the two sub-bands are well separated,
$\bar\varepsilon_2-\bar\varepsilon_1\gg \hbar\omega_0$, and take $\bar\varepsilon_2=0$ as the reference 
energy level.
 
While more general (though standard) expressions are derived in the Supplemental Material,
here we only report the expressions obtained for $g_{11}=g_{12}=g_{21}=0$, so that the system is not 
SC until the carrier density reaches the value such that $\mu=0$. For $0<\mu<\hbar\omega_0$, $T_c$ 
increases with increasing $\mu$
\[
T_c\approx 1.14\,\sqrt{\hbar\omega_0\mu}\,{\mathrm e}^{-1/\lambda_{2}},
\]
where we merged into the single dimensionless coupling $\lambda_2\equiv N^0_2g_{22}$
the intraband $g_{22}$ coupling and the density of states $N^0_2$ of the $\ell=2$ sub-band.
For $\mu>\hbar\omega_0$, $T_c$ saturates to its maximum (BCS) value
$T_c^{max}\approx 1.14\,\hbar\omega_0\,{\mathrm e}^{-1/\lambda_2}$.
Thus, $T_c(\mu)=0$, for $\mu<0$,
$T_c(\mu)=T_c^{max}\sqrt{\mu/\hbar\omega_0}$, for $0\le\mu\le\hbar\omega_0$, 
$T_c(\mu)=T_c^{max}$, for $\mu\ge\hbar\omega_0$, and
the range of variation of $\mu$ which corresponds to an increasing $T_c$ is a direct measure 
of the characteristic energy scale of the pairing mediator, $\hbar\omega_0$. 

We now proceed to fit the curve $\overline T_c(V_g)$ extracted from the experimental data by means of
EMT, with our theoretical curve $T_c(\mu)$. The relation between $V_g$ and $\mu$ is only approximately linear
and its determination is described in detail in the Supplemental Material. 
The resulting critical temperature for various values of $V_g$ is reported in Fig. \ref{fig-Tc}
(blue dashed line and crosses). From the fit, we finally obtain the dimensionless 
coupling constant $\lambda_2\approx 0.25$ (which, consistently with our assumption, is in the weak coupling regime) 
and $\hbar\omega_0\approx 23$\,meV, which  is the energy  typical of phonons in STO\cite{koonce}.

In conclusion, we describe the superconductivity in LAO/STO and LTO/STO oxide interfaces within a scenario in which
SC puddles form a percolating network. In this framework we showed here that the sheet resistance in 
LTO/STO interfaces is very well described by EMT or by a Christmas tree model of RRN for an inhomogeneous 
2DEG with a substantial filamentary character. Fitting the experiments allows to extract the random distribution of $T_c$ at various carrier densities (or $V_g$). 
A similar approach is also adopted to fit the micrometrically averaged superfluid density in LAO/STO \cite{bert2012}.
Assuming an effective two-band model with superconductivity triggered by the presence of few high-mobility carriers 
in a higher band, we account for the density dependence of the intra-puddle $T_c$ within a simple BCS weak coupling scheme. 
As an important by-product we find that the range of variation in $V_g$ of the (average) intra-puddle $T_c$ is directly related (via $\mu$)
to the typical energy of the pairing mediator $\omega_0$. 

S.C. and M.G. acknowledge financial support from ``University Research Project'' of the ``Sapienza''
University n. C26A115HTN. This work has been supported by the Region Ile-de-France in the framework of CNano IdF and Sesame programs, 
and by the DGA PhD program. Research in India was funded by the Department of Information Technology, Government of India.

\eject
\newpage

\appendix
\begin{center}
{ \bf {\Large  Supplemental Material for
Inhomogeneity and multiband superconductivity at the LaTiO$_3$/SrTiO$_3$ and  LaAlO$_3$/SrTiO$_3$ interfaces}}
\end{center}

In this supplemental material we provide further detail about the following two aspects of our work: the 
random resistor network used to model the inhomogeneous character of the oxide interfaces and to fit the 
experimental resistivity curves (section A) and the two-band model used to describe superconductivity 
inside the superconducting (SC) puddles (section B).

\section{A. $\:$ Details on the Christmas-tree model}
\begin{figure}
\centering
\begin{minipage}{0.5\textwidth}
	\centering
	\includegraphics[width=0.8\linewidth]{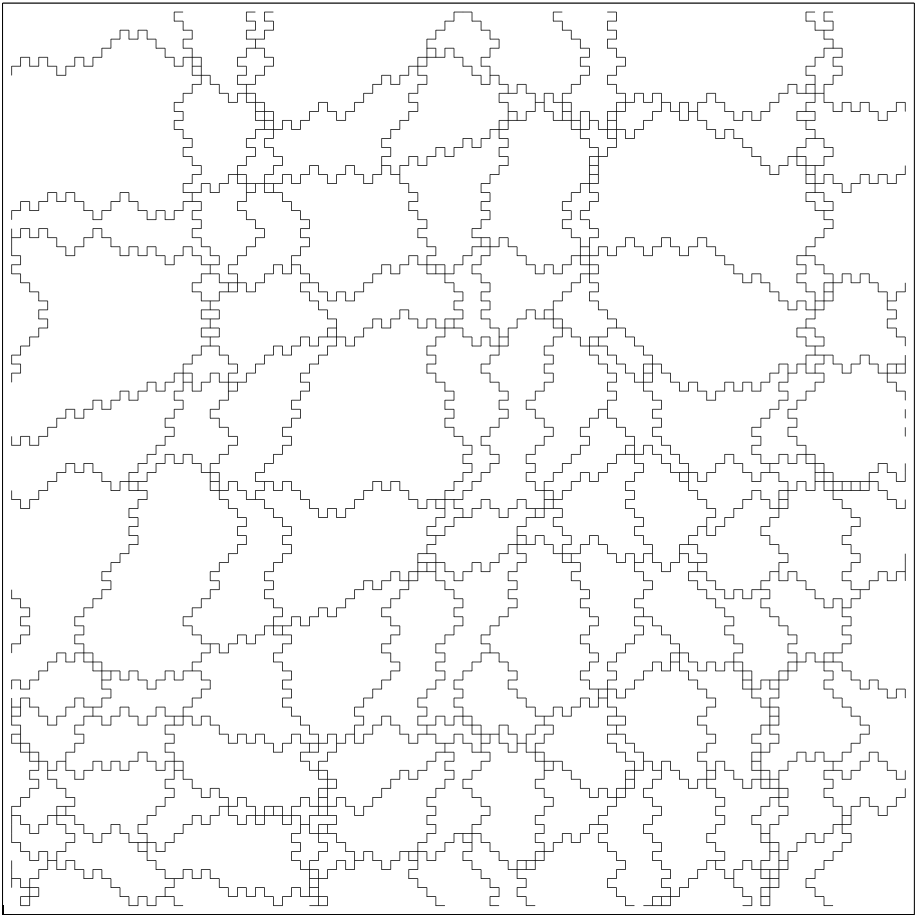}
	\caption{Symmetrized random walk cluster without patches; weight $w=0.2$.}
	\label{Cluster_SRWnP_paper}
\end{minipage}%
\begin{minipage}{0.5\textwidth}
	\centering
	\includegraphics[width=0.8\linewidth]{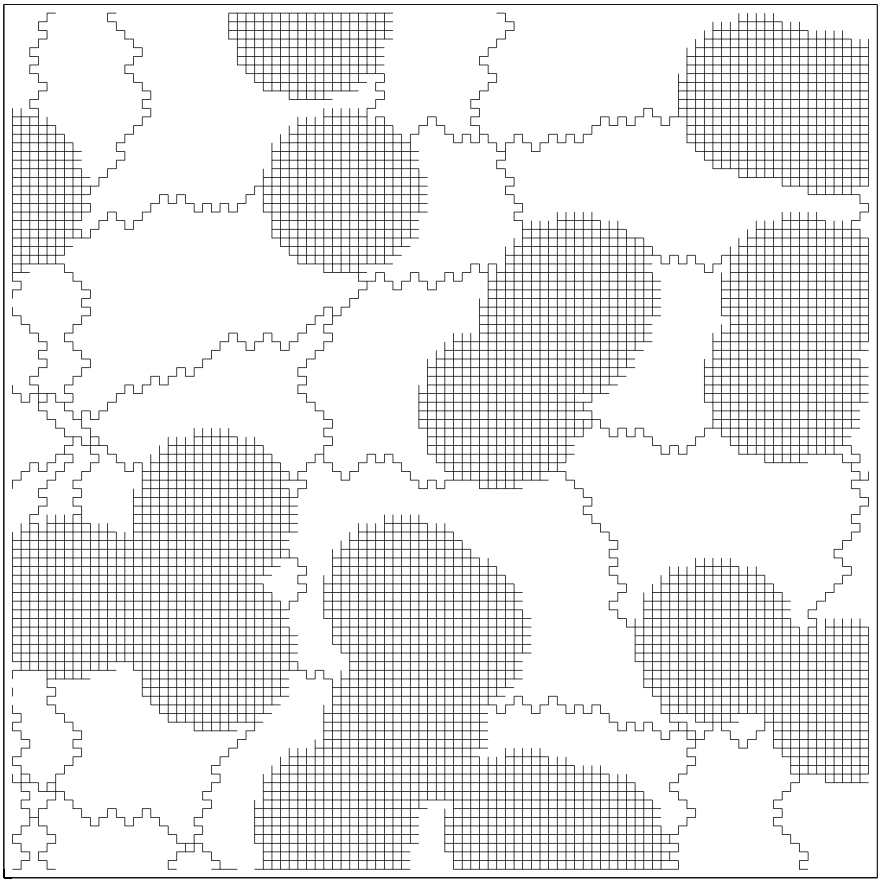}
	\caption{Symmetrized random walk cluster with patches; weight $w=0.5$.}
	\label{Cluster_SRWP_paper}
\end{minipage}
\centering
\begin{minipage}{0.5\textwidth}
	\centering
	\includegraphics[width=0.8\linewidth]{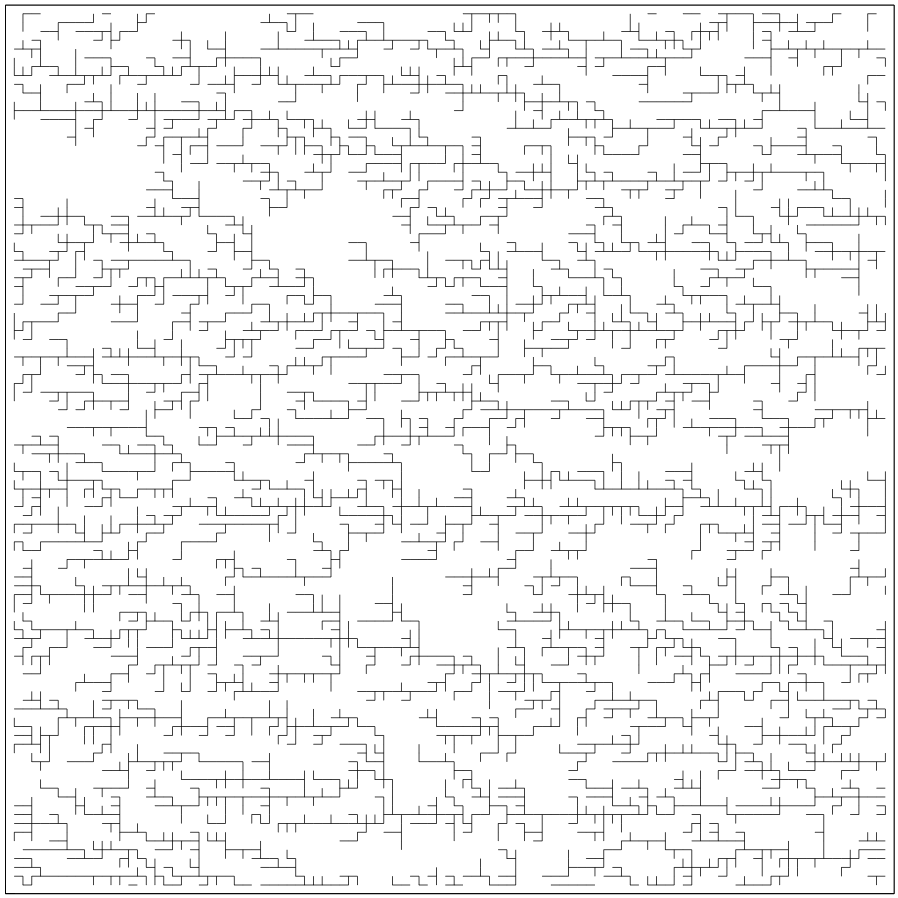}
	\caption{Diffusion limited aggregation cluster without patches; weight $w\approx0.3$.}
	\label{Cluster_DLAnP}
\end{minipage}%
\begin{minipage}{0.5\textwidth}
	\centering
	\includegraphics[width=0.8\linewidth]{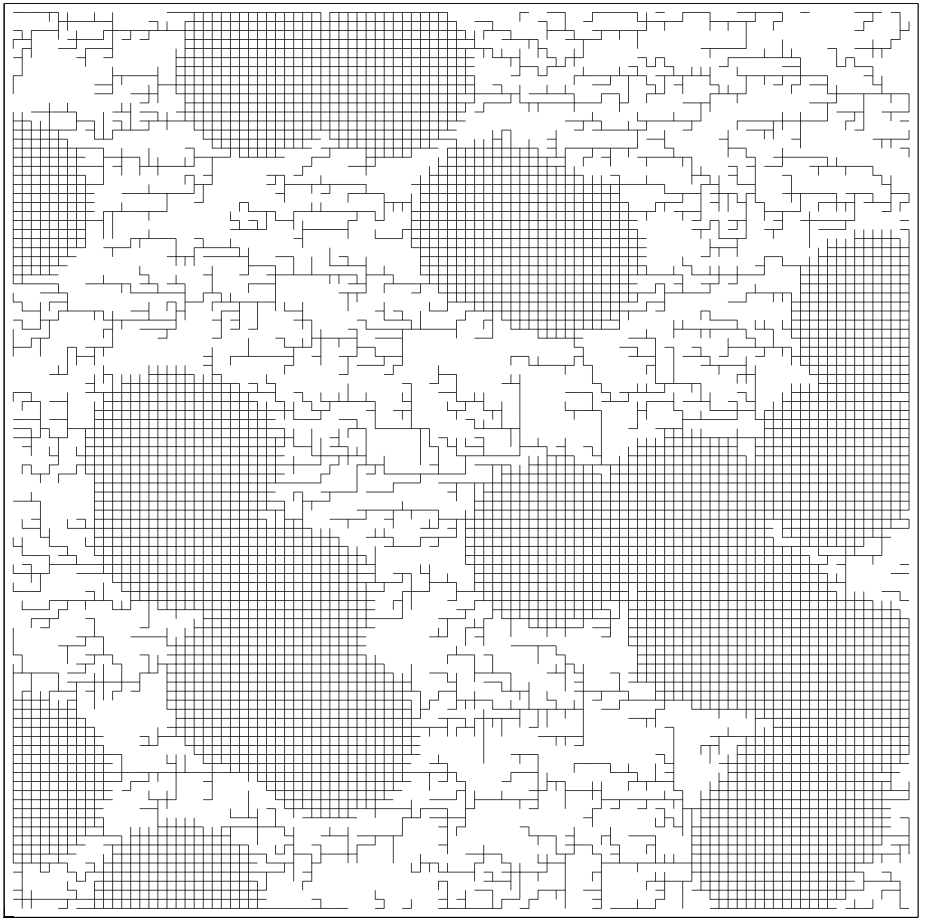}
	\caption{Diffusion limited aggregation cluster with patches; weight $w=0.6$.}
	\label{Cluster_ws06.pdf}
\end{minipage}
\end{figure}

To describe the inhomogeneous electric conductivity of LaTiO$_3$/SrTiO$_3$ and  LaAlO$_3$/SrTiO$_3$ interfaces,
we solve a random resistor network (RRN) consisting of two types of bonds: (i) normal bonds with a constant finite 
resistance $R_0=1$ and (ii) SC bonds with a temperature-dependent resistance 
$R^i=R_0\,\theta(T-T^i_c)$, where $T^i_c$ is the critical temperature of bond $i$ and $\theta(x)$ is the Heaviside 
step function. The critical temperatures are random variables obeying a Gaussian distribution.
The SC bonds form a spatially correlated cluster generated through a growth process known as 
\emph{diffusion limited aggregation} (DLA) on which patches of circular shape with a diameter about $30$ bonds long 
are superimposed, leading to a structure as shown in figure \ref{Cluster_ws06.pdf} (for details on the cluster growth 
see Ref. \onlinecite{BCCG}). The number of patches is chosen to yield RNNs with fractions of SC to total bonds equal to 
$w=0.3, 0.4, 0.45, 0.5, 0.55, 0.6, 0.65, 0.7$. 

Solving numerically Ohm's law on the bonds and  Kirchoff's law for current conservation at the nodes of the network, 
the resistivity curves $R_w(T)$ are obtained. For a better statistic, we calculate $R_w(T)$ for $5$ different 
realizations for each weight $w$ and take the arithmetic average. These curves are then used to fit the experimental 
curves $R_V$ measured at voltages $V=50, 70, 100, 150, 200$\:[V] in the following way: The high-temperature sheet 
resistance of the numerical solution is rescaled to the experimental one, with the exception of $100$[V], $150$[V] 
and $200$[V], where, to avoid the high temperature step, the fit is only done for the temperature interval 
$[0.03, 0.21]$ and the sheet resistance is rescaled to $R_V(T=0.21)$.  
The numerical temperature is then redefined to $\tilde{T}=(T-a)/b$, 
where $a$ and $b$ are chosen so as to minimize the difference 
\begin{equation}
d=\sum_{T_j}|R_V(T_j)-R_w(T_j)|,
\end{equation}
the sum running over all measured temperatures.
The rescaled curve which minimizes the difference $d$, $R_{w=\bar{w}}(\tilde{T})$, is taken as the ``correct'' one and 
the sample at the corresponding voltage is assumed to have a weight of SC bonds equal to $\bar{w}$.
\begin{figure}[h]
	\includegraphics[width=1\linewidth]{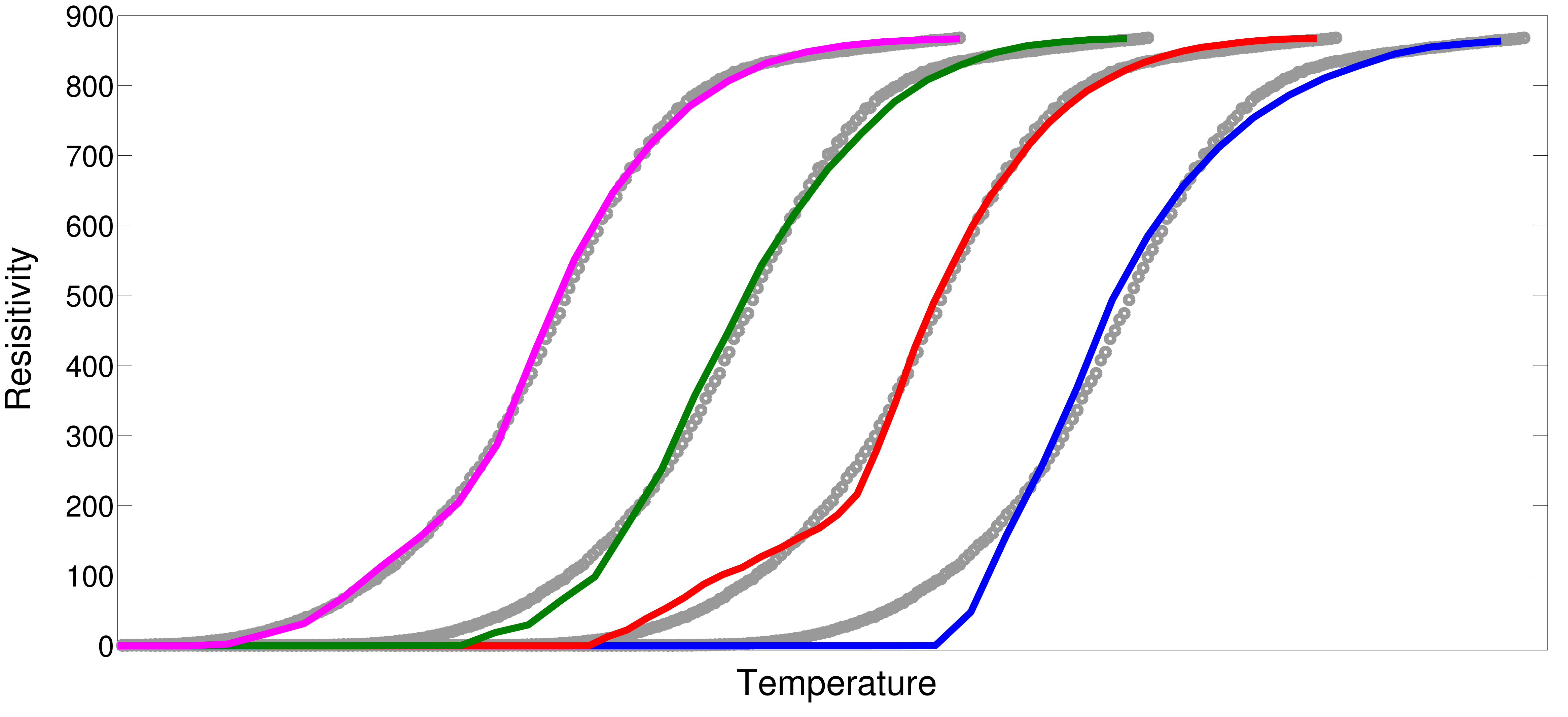}
	\caption{Fit of resistivity curve measured at $V=100$\;[V]. Grey curves report the experimental data $R_{V=100}$ 
	(duplicated and shifted for ease of reading), coloured curves report (from left to right) numerical calculations 
	obtained for DLA with patches, DLA without patches, SRW with patches, SRW without patches.}
	\label{R_T_V100_FIT}
\end{figure}

In addition to the quasi 1D long-range connectivity and quasi 2D short-range connectivity, the Christmas-tree model 
has the characteristic of closed loops covering almost entirely the cluster. With the size of a bond about $10\,nm$, 
these loops are of the order of a micrometer. This aspect is relevant when interpreting superfluid density
measurements. Indeed, the question arises about the magnetic response measured when performing a SQUID experiment, 
the SQUID pick-up coil recording the local magnetic field produced by a field coil that extends over a 
few micrometers. Under these conditions, the measurement resolution exceeds the (average) size of a loop and the 
measurement cannot distinguish whether a closed loop encompasses a SC or a non-SC 
region. More precisely, the SC loops being smaller than the experimental resolution, the SQUID sees 
all bonds inside a loop as if they were SC. As a result, the SQUID experiment can yield a superfluid 
fraction $w\sim1$ all the while the transport measurements yield resistivity curves that can be fitted with 
$\bar{w}\sim0.5$.

While fitting the experimental data may seem to require simply a cluster with a small long-range connectivity and 
a high density, we stress that the issue is more subtle than that. Indeed, we considered a series of clusters 
sharing some characteristics of the Christmas-tree model and calculated the corresponding resistivity curves (see 
figure \ref{R_T_V100_FIT}). Starting from the assumption that the cluster should be filamentary, dense and 
possibly exhibit loops, we considered a symmetric random walk (SRW) as shown in figure \ref{Cluster_SRWnP_paper}.
Even though the connectivity is rather small, it is too large to yield pronounced tails as experimentally 
observed. In order to make the system more 1D and possibly denser, we reduced the number of random walks and 
superimposed SC patches (obtaining a cluster similar to the Christmas-tree) as shown in figure \ref{Cluster_SRWP_paper}.
One does obtain a tailish feature at low temperature, which however is not in agreement with the experimental data.
Moving on to a DLA cluster without patches (figure \ref{Cluster_DLAnP}), we found resistivity curves leading to 
rather satisfactory fits of the transport data. However, this cluster does not exhibit loops and, as a theoretical 
drawback, one cannot easily tune the density with keeping the long-range connectivity low.
Finally, when adding patches, these two limitations are overcome, which leads us to conclude 
that our Christmas-tree model captures the main features of the SC state at the interface.

\section{B. $\:$ Two-band superconductivity for LXO/STO interfaces}
We propose that the 2DEG formed at the LTO/STO interface may be described as 
a multiband system, and the occurrence of superconductivity may be captured by means 
of a BCS-like Hamiltonian 
\begin{eqnarray}
\mathcal H &=&\sum_{\mathbf{k},\ell}
\sum_{\sigma=\uparrow,\downarrow}\left(\varepsilon_{\mathbf{k},\ell}-\mu\right)
c^\dagger_{\mathbf{k},\ell,\sigma}c^{\phantom\dagger}_{\mathbf{k},\ell,\sigma}
\nonumber\\
&+&\frac{1}{N}\widetilde{\sum_{\mathbf{k},\mathbf{k}'\atop \ell,\ell'}}g_{\ell\ell'}\,
c^\dagger_{\mathbf{k},\ell,\uparrow}c^\dagger_{-\mathbf{k},\ell,\downarrow}
c^{\phantom\dagger}_{-\mathbf{k}',\ell',\downarrow}
c^{\phantom\dagger}_{\mathbf{k}',\ell',\uparrow}\label{eqbcs}
\end{eqnarray}
where $c^\dagger_{\mathbf{k},\ell,\sigma}
(c^{\phantom\dagger}_{\mathbf{k},\ell,\sigma})$ creates (annihilates) an electron
with two-dimensional wave vector $\mathbf{k}$, parallel to the plane of the
interface, and spin projection $\sigma$, belonging to the $\ell$-th sub-band, 
with dispersion law
\[
\varepsilon_{\mathbf{k},\ell}=\bar\varepsilon_\ell+\frac{\hbar^2{k_x}^2}{2 m_{\ell,x}}+\frac{\hbar^2{k_y}^2}{2 m_{\ell,y}}.
\]
Hereafter, $\bar\varepsilon_\ell$ label the levels arising both from the level quantization inside the
confining potential well and from the various bands mostly due to the $t_{2g}$
orbitals of Ti in STO. $m$ is the (possibly anisotropic)  effective mass of the 2D
charge carriers, $\mu$ is the chemical potential, $N$ is the number of 
$\mathbf{k}$ vectors allowed by the boundary conditions in the first Brillouin zone 
of the two-dimensional lattice. The parameters $g_{\ell\ell'}$ are the 
(intraband for $\ell=\ell'$, interband for $\ell\neq\ell'$)
pairing amplitudes, and the tilde in Eq. (\ref{eqbcs}) indicates that the sums are
restricted to electron states such that $|\varepsilon_{\mathbf{k},\ell}-\mu|,
|\varepsilon_{\mathbf{k}',\ell'}-\mu|\le \hbar\omega_0$, where $\omega_0$ is a
characteristic frequency of the mode that mediates pairing.

In our description of the LTO/STO interface, we assume, for simplicity, 
that only two sub-bands, labeled by $\ell=1,2$, are involved. The $\ell=1$ sub-band 
contains the low-mobility carriers found in Refs. \onlinecite{espci2,bell}, whereas the $\ell=2$
sub-band  accommodates the high-mobility carriers. The SC  puddles are
regions where the sub-band $\ell=2$ is locally filled, whereas the metallic (or weakly localized)
background corresponds to regions where the sub-band $\ell=2$ is empty. 
To reproduce the experimental results, we are led to assume that 
$g_{11}\ll g_{12},g_{21}\ll g_{22}$ (this condition is consistent with
the analysis of two-band model in Ref. \onlinecite{balatsky}, but different assumptions
 $g_{11}\lesssim g_{12},g_{21}$ or $g_{12},g_{21}\lesssim g_{22}$, would lead to similar
 conclusions as far as the present analysis is concerned). 
Finite $g_{12},g_{21}$ guarantee that superconductivity is induced
in the first sub-band, as soon as it establishes in the second sub-band.
 
The equation that determines the SC critical temperature $T_c$
in a two-band system is
\[
\left[1-g_{11}\Pi_{1}(T_c)\right]\left[1-g_{22}\Pi_{2}(T_c)\right]=
g_{12}g_{21}\Pi_{1}(T_c)\Pi_{2}(T_c)
\]
where
\[
\Pi_{\ell}(T)=\int_{|z|<\beta\hbar\omega_0}\frac{dz}{2z}N_{\ell}(z)
\tanh\left(\frac{z}{2}\right),
\]
$\beta=(\kappa_B T)^{-1}$, 
$N_\ell(z)=N_\ell^0\vartheta(z-\beta\bar\varepsilon_\ell+\beta\mu)$ is the 
density of states of the $\ell$-th sub-band (in the dimensionless
variable $z=\beta\varepsilon$), with $N_\ell^0=a^2 \sqrt{m_{\ell,x}m_{\ell,y}}/(2\pi\hbar^2)$, 
$a$ being the spacing of the two-dimensional lattice. There are
four possible regimes for $\Pi_{\ell}(T)$: if $\bar\varepsilon_\ell-\mu>\hbar\omega_0$,
the chemical potential falls below the bottom of the band and there are no states
available for pairing within the shell of width $2\hbar\omega_0$, hence
$\Pi_{\ell}(T)=0$; if $0<\bar\varepsilon_\ell-\mu<\hbar\omega_0$, the chemical 
potential still falls below the bottom of the band, but there are some states
available for pairing within the shell of width $2\hbar\omega_0$, hence
$\Pi_{\ell}(T)=N_\ell^0\ln[\sqrt{\hbar\omega_0/(\bar\varepsilon_\ell-\mu)}]$; if 
$-\hbar\omega_0<\bar\varepsilon_\ell-\mu<0$, 
the chemical potential falls within the band, although there are some states
unavailable for paring within the shell of width $2\hbar\omega_0$, hence
$\Pi_{\ell}(T)=N_\ell^0\ln[1.14\,\beta\sqrt{(\mu-\bar\varepsilon_\ell)\hbar\omega_0}]$;
if $\bar\varepsilon_\ell-\mu<-\hbar\omega_0$, the chemical potential falls within 
the band and all the states within the shell of width $2\hbar\omega_0$ are available
for pairing. In this case the standard BCS result 
$\Pi_{\ell}(T)=N_\ell^0\ln(1.14\,\beta\hbar\omega_0)$
is recovered. We do not report here the crossover 
expressions when $|\bar\varepsilon_\ell-\mu|\lesssim \kappa_B T$; we also assumed
that $\hbar\omega_0\gg\kappa_B T$, since we shall show that our system is indeed in the
weak-coupling regime (i.e., $N_\ell^0 g_{\ell\ell'}\ll 1$ for all the coupling constants $g_{\ell\ell'}$).

We assume henceforth that the bottoms of the two sub-bands are well separated,
$\bar\varepsilon_2-\bar\varepsilon_1\gg \hbar\omega_0$,
and take $\bar\varepsilon_2=0$ as the reference energy level. For $g_{11}=0$,
the system is not SC until the filling reaches the value such that 
$\mu=-\hbar\omega_0$. However, $T_c$ will stay exponentially
small, until $\mu=0$. In particular, letting $\mu\equiv-\hbar\omega_0+\delta\mu$, for
$\delta\mu\to 0^+$ we find
\[
T_c\approx 1.14\,\hbar\omega_0\,{\mathrm e}^{-2\hbar\omega_0/({N^0_1N^0_2}g_{12}g_{21}\delta\mu)}.
\]
In the following, we assume that $T_c=0$ in this regime.
In the range $0<\mu<\hbar\omega_0$, $T_c$ becomes sizable and increases with increasing $\mu$.
Neglecting small corrections due to $g_{12}$ and $g_{21}$ one finds
\begin{equation}
T_c\approx 1.14\,\sqrt{\hbar\omega_0\mu}\,{\mathrm e}^{-1/(N^0_2g_{22})},
\end{equation}
while, for $\mu>\hbar\omega_0$, $T_c$ saturates to its maximum (BCS) value
\begin{equation}
T_c^{max}\approx 1.14\,\hbar\omega_0\,{\mathrm e}^{-1/(N^0_2g_{22})}.
\end{equation}
Thus, $T_c(\mu)=0$, for $\mu<0$,
$T_c(\mu)=T_c^{max}\sqrt{\mu/\hbar\omega_0}$, for $0\le\mu\le\hbar\omega_0$, 
$T_c(\mu)=T_c^{max}$, for $\mu\ge\hbar\omega_0$, and
the range of variation of $\mu$ which corresponds to an increasing $T_c$ is a direct measure 
of the characteristic energy scale of the pairing mediator, $\hbar\omega_0$. 

The chemical potential $\mu$ is related to the variation of the carrier density $\delta n$ in the second 
sub-band as $\mu=\delta n/(2 N^0_2)$. On the other hand, we extract the value of $\delta n(V_g)$ from a 
self-consistent calculation of the sub-band structure in the quantum confining potential \cite{espci2}, in 
the presence of the gate voltage $V_g$. The procedure is the following: for each $V_g$ the full numerical
solution of the coupled Schr\"odinger and Poisson equations yields the sub-band structure in the confining
potential; the Fermi level $E_F$ is then fixed to accommodate all the carriers; the SC sub-band is 
identified as the one that starts to be filled in correspondence of the value of $V_g$ at which superconductivity 
is observed to occur; the chemical potential $\mu$ entering in our two band model is obtained as the difference
between $E_F$ and the bottom of the SC sub-band.

We are thus able to extract the dependence $\mu(V_g)$ to describe the chemical 
potential entering the $\ell=2$ band of the SC carriers and obtain the fit to the experimental
data shown in Fig. 3, for the set of parameters considered in Ref. \cite{espci2}, where the sub-bands 
have all $d_{xy}$ character, with light isotropic mass $m_{x,y}\approx 0.7\,m_0$.


\begin{thebibliography}{99}
\bibitem{notafractal} We point out that here and in Ref. \onlinecite{BCCG} the fractal character 
(of the Diffusion Limited Aggregation type) of the 
generated clusters has no intentional physical meaning, but is merely a technical artifice to generate dense, and 
yet filamentary, spatially correlated structures needed to reproduce tailish behavior of the
the sheet resistance curves of oxide interfaces.
\end{thebibliography}
\end{document}